# Enhancing operations management through smart sensors: measuring and improving well-being, interaction and performance of logistics workers

Aloini, D., Fronzetti Colladon, A., Gloor, P., Guerrazzi, E., & Stefanini, A.

# Enhancing Operations Management through Smart Sensors: Measuring and Improving Well-Being, Interaction and Performance of Logistics Workers

**Abstract**

**Purpose** – Human and environmental factors play an important role in operations management activities since they significantly influence employees' performance, well-being, and safety. Surprisingly, empirical studies about the impact of such aspects on logistics operations are still very limited. Trying to fill this gap, this research empirically explores human and environmental factors affecting the performance of logistics workers exploiting smart tools.

**Design/methodology/approach** – This research conducted an exploratory investigation of the material handling activities of an Italian logistics hub. Wearable sensors and other smart tools were used for collecting human and environmental features during working activities. These factors were correlated with workers' performance and well-being.

**Findings** – Results suggest that human attitudes, interactions, emotions, and environmental conditions remarkably influence workers' performance and well-being, however showing different relationships depending on individual characteristics of each worker.

**Implications** – Our research opens up new avenues for profiling employees and adopting an individualized human resource management, providing managers with an operational system capable to potentially check and improve workers' well-being and performance.

**Originality** – The originality of this study comes from the in-depth exploration of human and environmental factors using body-worn sensors during work activities, by recording individual, collaborative, and environmental data in real-time. To the best of our knowledge, this is the first time that such a detailed analysis has been carried out in real-world logistics operations.

*"Great things in business are never done by one person. They're done by a team of people."* -- Steve Jobs

# 1. Introduction

In many business sectors, warehouse management is a key element of a company's overall logistics process, as it determines the ability of the company to deliver products to customers on demand (Grosse et al., 2017). In addition, it also impacts the company's finances, given the costs associated with managing the inventory. Recognized as a fundamental aspect of many organizations, the efficiency and effectiveness of warehouse management have attracted significant attention by both scholars and practitioners (Richards & Grinsted, 2020).

In particular, motivated by the rise of Industry 4.0, numerous novel technologies were introduced in this field, such as Automated Guided Vehicles (AGV), robots, smart tools, Augmented Reality (AR), and Virtual Reality (VR). The recent emergence of virtual technologies – VR and AR – appears to be especially promising for supporting logistics activities, but is still quite unexplored (Reif & Walch, 2008; Jovanovikj & Engels, 2020; Zywicki & Bun, 2021). AR can be successfully used for improving logistics workers' efficiency thanks to visual guidance (e.g., during order picking), while VR can be helpful for managing remote objects in dangerous environments and for training workers (Fang et al., 2019; Matsumoto, 2019; Fang & An, 2021; Yigitbas et al., 2020; Grabowski et al., 2021).

Although automation and pervasiveness of digital technologies in logistics activities increased significantly recently, warehouses still heavily rely on human operators. Indeed, human workers guarantee the flexibility required by material handling processes thanks to their cognitive and physical abilities, which are still hard to replicate with automated systems in a cost-effective way (Grosse et al., 2016). Consequently, workers' attitude and behaviors, and their collaboration dynamics, are among the fundamental drivers of system performance – in term of efficacy (e.g., mistakes in order picking) and efficiency (e.g., times of material handling), as well as workplace safety and individual wellness (Grosse et al., 2015).

These reasons reinforce the need to consider in more detail human and work environment factors to increase business performance, employees' safety, and their well-being. Surprisingly, empirical studies exploring the impact on logistics operations of these factors are still very limited (Grosse et al., 2017). Indeed, behavioral dynamics and environmental factors have started to be investigated in logistics only recently, as part of the Behavioral Operations Management (BOM) approach (Loch and Wu, 2007; Gino & Pisano 2008; Croson et al., 2013; Fahimnia et al., 2019).

The lack of strong evidence of the influence of human and work environment aspects in logistics has been recently noted (Neumann et al., 2021) and appears to be due to the methodological difficulty of studying behavioral elements in real-life contexts. Measuring and evaluating individual and collaborative behaviors during work activities is indeed a challenging task. For example, sources of fatigue, injury and work inefficiency that result from badly designed work systems in logistics activities are often not well understood (Grosse et al., 2016).

However, the recent wide availability of wearable sensors and similar smart tools is providing researchers with the opportunity to collect and analyze data about human-behavior factors and work environment conditions, through effective data-driven methodologies (Chaffin et al., 2017; Ni et al., 2020; Stefanini et al., 2020). These new sensor-based tools (e.g., smartphones, smartwatches, and sociometric badges) allow researchers to quantitatively study the behaviors of logistics employees and their environmental conditions in deeper and more efficient ways, directly and without compromising their operational activities.

To shed some light on this issue, this work aims at empirically exploring human and environmental factors that affect the performance of warehouse workers during logistics activities, by using an innovative investigation approach.

Specifically, this research tries to answer to the following research questions (RQs):

(RQ1): Which are the most important individual, collaborative, and environmental factors affecting workers' well-being and productivity during logistics activities?

(RQ2): How can wearable sensor tools support the investigation of individual, collaborative, and environmental factors in Operations Management?

To respond to the RQs above, an explorative study was carried out in a semi-automated logistics hub. We adopted an innovative body sensing approach, using smartwatches – which were equipped with a mobile app developed at the MIT Center for Collective Intelligence (Gloor et al., 2018) – capable of measuring variables such as heart rate, body movement, and interpersonal interactions. In addition, we recorded environmental variables such as humidity, temperature, and luminosity using additional appropriate smart devices.

The uniqueness of this study comes from the in-depth exploration of workers' behaviors and environmental conditions, by recording individual, collaborative, and environmental data in real-time. To the best of our knowledge, this is the first time that such a detailed analysis is carried out for real-world logistics operations. Our explorative results may help researchers and managers by highlighting relevant human and environmental factors that affect warehousing performance. These elements could be considered for planning logistics activities, designing warehouse structures, and training logistics staff.

Our study also contributes to the operations management methodology research (George et al., 2016). It represents an early attempt to systematically investigate the effect of human and environmental factors on performance during operational activities, exploiting wearable tools and smart environmental sensors. In this way, this research provides a first answer to the open issue of "how" to quantitatively investigate such factors in real business environments (Croson et al., 2013; Brocklesby, 2016). The approach described also opens up new avenues for adopting an individualized, "one-to-one", management style reducing the bias related to the different individual characteristics of workers (Baydar, 2002).

The paper is structured as follows: Section 2 overviews related studies on human and environmental factors in logistics. Section 3 describes the research method and reports the study measurements. Section 4 presents the results, while Section 5 discusses them and outlines the main theoretical and managerial contributions. Finally, Section 6 concludes the paper, considers the study limitations, and outlines directions for future research.

## 2. Background

Recent literature shows that new levers can be adopted by companies to enhance their performance; particularly the adoption of corporate social responsibility (CSR) and environmental management system (EMS) supports employee satisfaction and has positive externalities on society (Ikram, Sroufe, et al., 2019; Ikram, Zhou, et al., 2019). Recognizing the high relevance of warehousing – and more broadly, logistics activities for business performance – past research has extensively addressed this topic from multiple perspectives (e.g. Masae et al., 2020; Richards & Grinsted, 2020). In the last decades, researchers have developed both conceptual and analytical models to improve warehouse operations and increase the efficiency of logistics systems, particularly in the field of order picking that is acknowledged as one of the most impactful (Tompkins et al., 2010; Grosse et al., 2017). In doing so, scholars have suggested and tested different solutions for improving warehouse design and management, e.g. layouts, vehicles and operators routing, item location choices, picking order logic, advanced information systems to measure and monitor performances, etc. (e.g., Shiau and Lee, 2010; Masae et al., 2020). More recently, VR and AR technologies are starting to gain ground in industrial applications, for example they have been adopted in logistics for improving order picking performance (Fang & An, 2021; Jovanovikj & Engels, 2020; Matsumoto, 2019; Yigitbas et al., 2020) and to support the effective management of operations in the warehouse (Mourtzis et al., 2018). Their applicability has proven to be useful for providing positioning instructions, both for maps navigation but also in large warehouse facilities for inventory management and package retrieval, assisting the operators to efficiently manage large facilities with changing stocks (such as this is the case in logistics warehouses) (Fraga-Lamas et al., 2018).

However, to the best of our knowledge, few studies have paid attention to human factors, which – despite the pervasive automation and digitalization trend, known as "Industry 4.0" (Masae et al., 2020; Winkelhaus & Grosse, 2020) – still have a large impact on system performance (Grosse et al., 2015; Calzavara et al., 2018). Indeed, working conditions have been shown to have a remarkable influence on job satisfaction (Awan et al., 2013). Many human aspects appear to have a strong impact on warehouse system performance, such as physical fatigue, individual attitudes and psychological traits, collaboration dynamics between workers, individual and group motivation, and personal inclination to learning (Glock et al., 2019; Tucker et al., 2018; Ellinger et al., 2005; Jordan et al., 2019; Masae et al., 2020). Neglecting these aspects is widely recognized as a big limitation of current research (Grosse et al 2017). Therefore, an open call for future research focusing on the link between warehousing system design and human aspects exists in the literature (Neumann & Dul, 2010; Battini et al., 2017; Calzavara et al., 2017).

Beyond human behaviors, environmental factors seem also able to influence the operators' well-being and performance. Warehouses are complex constructs where environmental conditions play an important role, as they affect the overall ergonomics of the workplace. Past research showed that environmental variables such as temperature, humidity, air quality, and luminosity highly influence the working environment and can have a relevant impact on workers' safety, well-being, and performance in various contexts (Leather et al., 2003; Salvendy, 2012; Marqueze et al., 2015; Meister, 2018; Molka-Danielsen et al., 2018). Moreover, recent studies showed that higher attention to environmental factors (such as reducing waste) has positive externalities for the industry itself (Zimon et al., 2021), as well as CSR (Ikram, Sroufe, et al., 2019) – and the adoption of EMS (Ikram, Zhou, et al., 2019) are likely to have a positive impact on employees' quality of life. Again, scientific literature currently is lacking empirical evidence on the impact of environmental factors on logistics workers and their performance (Grosse et al., 2017).

In other words, research on the role of human and environmental variables in logistics activities is underdeveloped, even though it appears to be highly relevant for the improvement of system performance. A better understanding of such factors, indeed, may allow researchers a more reliable modelling of logistics systems to achieve higher overall performance and, at the same time, improve the safety and well-being of workers (Neumann & Dul, 2010; Battini et al., 2017; Calzavara et al., 2017). Particularly, even if there are some general indications that support optimal work climate, not all people respond in the same way to different factors (Lamberti et al., 2020). Consequently, to foster employee's well-being, managers are required to act in a customized-management style perspective considering, when possible, the individual differences of workers. Accordingly, this work tries to address this research gap through an exploratory empirical investigation.

In the following subsections, individual and collective human factors (Section 2.1), as well as environmental factors (Section 2.2), are discussed.

### *2.1 Individual and collective human factors*

The study of human factors can be defined as: *"the scientific discipline concerned with the understanding of interactions among humans and other elements of a system … in order to optimize human well-being and overall system performance."* (IEA Council, 2000). Though practitioners have long recognized the importance of humans in operations (Neumann and Dul 2010; Stefanini et al., 2020), and Operations Management textbooks have included sections on human factors for a long time (e.g. Heizer & Render, 2007; Wild, 1995), only recently has the literature focused on human factors in material handling activities for the improvement of system efficiency (Grosse et al., 2015).

Fatigue is the most investigated individual factor and clearly has an influence on material handling performance (Calzavara et al., 2017). It is defined both in psychological and physiological terms (Gawron et al., 2001) as a response to stress that can trigger mental and/or physical reactions within the human body (Schlick et al., 2010). In the first case, it is supposed to reduce performance when tasks require alertness and memory promptness or reactiveness, e.g., for information retrieval and processing. The second aspect includes a reduction in the capacity to perform physical work because of a previous physical effort. Both types of fatigue are known to negatively influence performance.

Focusing on this research area, Winkelhaus *et al.* (2018) investigated the influence of fatigue and learning on material handling performance and found a reverse U-shaped relationship between physical fatigue and cognitive performance. Moderate physical movement can improve information processing, while too much physical movement can reduce performance. Authors also underline that this is highly dependent on individual characteristics and call for more empirical studies. Calzavara *et al.* (2018) proposed a heart rate monitor device to monitor the physical fatigue of operators to improve their performance in logistics activities. The researchers presented a personalized analysis of fatigue, as the changes in rate and duration of the activity imply different energy expenditures. Similarly, Liu and Zou (2018) investigated physical fatigue of the operators trying to reduce awkward postures (Battini et al., 2011). Specifically, they studied a new design for express delivery systems based on human factors engineering. Results show a significant improvement in work efficiency (about 30%), while reducing the staff's body effort by nearly 60%.

Although fatigue remains the most observed aspect, some studies have tried to investigate other individual attitudes and behaviors (Grosse et al., 2017). For instance, (De Vries et al., 2016) explored the role of individual differences in picking performance with various picking tools and methods, showing that some fundamental personality traits and the age of the picker play a significant role in predicting picking performance. Previous research showed that a considerable amount of the variability of performance resides within people (Dalal et al., 2009). Another example, Elbert *et al.* (2017) built a simulation model to quantify the effects of deviations from pre-specified routes in order picking on workers' performance. They confirmed that the individual attitude of workers can induce them to modify their routes, reducing their expected performance.

In addition to individual elements, workers' collaboration dynamics appear to play a very important role. Although teamwork is often easy to "informally" observe, it is difficult to describe and even more difficult to model and analyze. At a broad level, teamwork is the process through which team members collaborate to achieve team goals (Driskell *et al*, 2018). An adequate collaboration between team members is essential to achieve high team performance, which is reflected in the quality and quantity of tasks completed (Salas et al., 2005; Sun et al., 2017). Teamwork is also a necessary condition for innovation performance (Awan & Sroufe, 2021).

In general, a high-performing team is also capable to provide the best results through commitment to high-quality standards (van de Brake et al., 2018). For example, in the study of Savelsbergh *et al.* (2010), a survey of 22 teams identified the main factors influencing team performance. These findings give evidence to the importance of team leadership, goal clarity and team learning behaviors. A more recent work by Fan *et al.* (2017) used process mining of a collaboration system log to identify collaboration patterns leading to more efficient teamwork.

Even if collaboration dynamics appear to be important also for logistics activities, this aspect is largely neglected by the current literature (Grosse et al., 2017). Again, there is a need to consider individuality when discussing an employee's commitment to a team (Obembe, 2010) and to individual work performance (Othman & Mahmood, 2019).

In summary, given the preliminary evidence on the importance of fatigue, individual attitudes, and collaboration dynamics in logistics, our research contributes to this stream of literature through an exploratory study that takes into account the combined effects of these factors, using a data-driven approach.

*2.2 Environmental factors*

The physical characteristics of the work environment can have a direct impact on productivity, health, safety, well-being, concentration, job satisfaction and morale of the involved people, as such aspects affect the overall ergonomics of the workplace (Danna & Griffin, 1999; Leather et al., 2003; Sarode & Shirsath, 2014; Lan et al., 2014; Castaldo et al., 2018; Bellingan et al., 2020; Pigliautile et al., 2020; Kovalev et al., 2021).

Among factors that play an important role in the workplace, scholars include organization of spaces, temperature, shift timing, humidity, lighting, ventilation, noise, and air quality (e.g., Kralikova et al., 2019, Jin et al., 2016, Wijewardane & Jayasinghe, 2008; Lan et al., 2010; Marqueze et al., 2015; Wargocki & Wyon, 2017; McInnes et al., 2017). The relevance of each factor depends on the context under investigation. For instance, air quality matters in work environments where motorized vehicles emit significant amounts of CO2 or where there are many people in confined spaces (Leblebici, 2012; Molka-Danielsen et al., 2018), with both temperature and humidity being highly related to the perception of air quality (Fang & An, 2021). Noise can significantly affect work motivation and operations performance in workplaces where loud sounds are more frequent (Leather et al., 2003; Sloof & Van Praag, 2010; Akbari et al., 2013). A correct lighting level may improve human well-being, with industrial smart lighting systems providing benefits also in efficient work processes, reducing cost and risk of failure (Füchtenhans et al., 2021).

Broadly speaking, a more comfortable working environment seems to allow workers to better reach their potential and achieve superior performance (Lan et al., 2010).

The investigation of environmental factors in workplaces is therefore important since it allows an improvement of the work setting, which can ultimately impact business performance. Unfortunately, research on environmental factors in logistics, and specifically on warehouse activities, remains scarce. The lack of empirical evidence is partly related to the difficulty of evaluating the effects of environmental variables on workers, also considering the different needs of each operator based on personal aspects such as gender, age, and individual characteristics (e.g. Karjalainen 2012; McElroy & Morrow, 2010; Lan et al., 2010).

Recent research attempted to address this gap, mostly considering workplace ergonomics (Grosse et al., 2017). The advent of smart sensors is helping researchers and practitioners to directly monitor environmental and individual characteristics in real industrial settings (Oikonomou et al., 2016; Molka-Danielsen et al., 2018). For example, Molka-Danielsen and colleagues (2018) tried to assess temperature, humidity, and $CO_2$ concentration in order to provide safety advice at a logistics shipping base. Nevertheless, resulting evidence about the impact of environmental aspects on warehouse workers' performance are still inconclusive (Grosse et al., 2017).

Given the recognized relevance of factors such as temperature, luminosity, and humidity on performance in several business areas (e.g., Kamarulzaman et al., 2011; Lan et al., 2014;

Marqueze et al., 2015; McInnes et al., 2017; Al-Omari & Okasheh, 2017), we posit that a significant impact of such factors can be expected also in the context of a logistics hub.

Considering the literature we reviewed – showing the potential impact of individual, collaborative, and environmental factors on the well-being and productivity of workers – we present a study that addresses the above mentioned research gaps and integrates factors pertaining to multiple domains in a comprehensive model. Our effort is aligned with recent calls for future research that considered these aspects (Battini et al., 2017; Calzavara et al., 2017; Grosse et al., 2017).

## 3. Research Design

*3.1. Methodology*

This research adopts a novel data-driven approach to empirically investigate the effect of human and environmental factors on worker performance in material handling activities. Specifically, smartwatches – endowed with a mobile app devised by the Center for Collective Intelligence at the Massachusetts Institute of Technology (Gloor et al., 2018) – are employed to systematically measure operators' behavior-related variables – such as body movement, heart rate, voice, and operators' interactions – during working activities. We use smartwatch sensors because there are limitations in using only smartphone-embedded sensors (e.g., bias in heart rate measurement) (Gloor et al., 2018). Alternative technologies, such as sociometric badges, could offer powerful tools for gathering data on workplace interactions (including location and speech) (Dong et al., 2012). However, they are often more expensive and intrusive than smartwatches, making them relatively difficult to use in long-term studies.

Consistent with the methodology presented in past research (e.g., Jin et al., 2016), a smart thermo-hygrometer and a lux meter were deployed to continuously track work environment conditions, particularly humidity, temperature, and luminosity. Given the lack of strong empirical evidence on the topic and the novelty of our research approach, an exploratory case study (Yin, 2017) was carried out in a semi-automated logistics center of an Italian company.

The case study was conducted in the following four phases:

- <u>Context evaluation</u>. An analysis of the logistics hub context was conducted to prepare the research study and to preliminarily test the smart tools (i.e., smartwatches, mobile app, thermo-hygrometer, and lux-meter) and the related metrics. Specifically, a preliminary investigation of the warehouse allowed to define the research settings (e.g. workers and activities involved in the study), to set up and fine tune the smart sensors and tools, and to identify the variables to be considered.

- <u>Data collection</u>. Operator behavior and other human-related variables (e.g., operators' interactions, body movement, speaking, proximity, and heart rate) were collected during the work activities through the smartwatches and the mobile app, while thermo-hygrometer and lux-meter provided working environment conditions (i.e., temperature, humidity, and luminosity). Finally, the operational performance of workers, i.e. their hourly productivity, was retrieved from the Warehouse Management Systems (WMS).

- <u>Data pre-processing</u>. To achieve comparable metrics and create appropriate features for the subsequent analysis phase, data collected in the previous step were pre-processed and standardized. The variables used in this study are described in section 3.3. In addition to raw scores, we considered the one-hour lag of each predictor since some observable factors could have a deferred effect on performance.

- <u>Data analysis</u>. A machine learning regression model based on unbiased boosted trees with categorical features, namely CatBoost (Prokhorenkova et al., 2018), was trained to understand whether environmental conditions, body sensing parameters, and measures of collaboration dynamics could support the prediction of hourly productivity of workers. Linear models were not applied because the relationships of predictors with the dependent variable were not linear. Subsequently, SHapley Additive exPlanations (SHAP) were used to determine feature importance, for the predictions obtained through CatBoost (Lundberg & Lee, 2017; Lundberg et al., 2019).

Lastly, feedback from managers and staff was collected in focus groups to confirm the interpretation of the results and to develop additional implications from a managerial viewpoint.

*3.2. Case study and Data Collection*

The investigation was conducted in a logistics hub of one of the largest producers of tissue paper in Europe. The warehouse, located in the center-north area of Italy, extends for 24,000 square meters and is dedicated to the storage of paper for hygienic and domestic uses.

Operational activities of the warehouse are the stocking of inbound pallets coming from production plants and the picking of stocked pallets for customer delivery. Each pallet has standard dimensions (80 x 120 cm), while the weight is not critical for the handling operations because of the lightness of paper-tissue. Typical working shifts consist of 9 workers: 4 for stocking, 4 for picking, and 1 team leader. Order picking operators work from Monday to Friday, usually 8 hours a day in two shifts (morning and afternoon) that alternate on a weekly schedule. The presence of workers is still mandatory for the picking activities since the pallet shuttle only supports pallet movement inside the shelving, while employees move the pallets with forklifts to and from the loading areas.

The data collection was carried out from March to May 2019 on picking activities, since they are more complex than stocking activities and demand a greater coordination among workers (Grosse et al., 2017). The observed team consisted therefore of 5 workers, 4 picking operators and the team leader. A total of about 1600 hours of work were analyzed.

For collecting worker behaviors, the 5 workers of the picking team were equipped with smartwatches during their work activities for the whole investigation period. The tools permitted to measure and gather operators' behaviors in term of (i) physical movement/human activity, (ii) indoor localization, (iii) speech features, (iv) proximity to colleagues, and (v) interactions with colleagues. All personal data were anonymized, and the speech content was not recorded to be compliant with privacy law.

As regards the environmental aspects, thermo-hygrometer and lux-meter systems – with datalogger function – were synchronized and placed in the warehouse area to constantly monitor and collect data on working conditions.

Finally, data extracted by the WMS permitted the assessment of workers' productivity, i.e., the hourly number of processed loading units (i.e., pallets).

Figure 1 summarizes the data collected by the study and related sources.

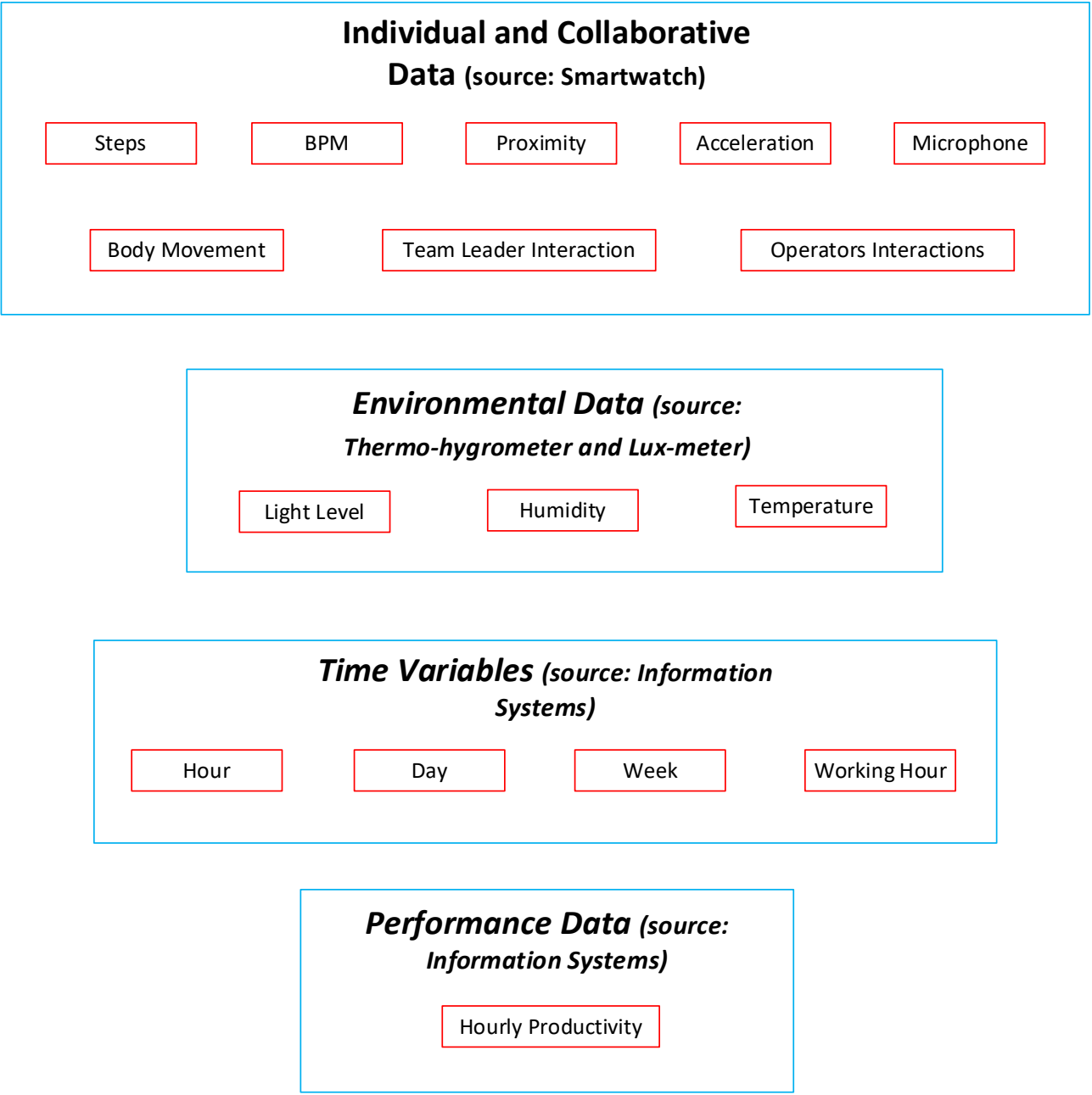

**Figure 1** – Data collected by the study and related sources

*3.3 Study Variables*

In this section, dependent and independent variables adopted in the study are described.

The *dependent variable* tracks the employee productivity in term of loading units, i.e. pallets, correctly processed per hour.

The list of *independent variables* is reported in Table 1, together with their descriptions. Measures were repeated every hour, and continuous measures averaged on the hour.

Table 1 - Description of the independent variables

| Variable | Description |
| --- | --- |
| **Userid** | Operator ID number. It is a number that allows to identify the single operator. |
| **Hour** | Hour of the day (range: 5-23). Evening shifts have proved to affect work performance, for example being related to sleepiness (Gillberg et al., 2003). |
| **Day** | This variable is "1" for Monday, "2" for Tuesday and so on, till "5" for Friday. A recent study has confirmed that across weekdays, employees' fatigue may vary (Elfering et al., 2021). |
| **Week** | Week number in the experiment (e.g.: "1" means the first week). Job performance is not a stable characteristic of certain employees, but varies from year to year, month to month, week to week (Vahle-hinz, 2021). |
| **Working_Hour** | Number of hours elapsed since the beginning of the shift. A decrease in performance efficiency due to drowsiness and lack of concentration can be expected over the workday (Rosa et al., 1985). |
| **Bpm** | Average value of the heart rate, measured in beats per minute (bpm) every minute. Heart rate is a proxy measure of fatigue (Shen, 2018) |
| **Max_Bpm** | Maximum value of the variable "Bpm" in one hour. BPM peak may suggest extra-effort or stress. Tachycardia may be indicative of mental stress, which in turn can decrease performance (Jenks et al., 2020). |
| **N_Bpm_Max** | Number of times (in one hour) in which the bpm exceeded the average bpm value by at least one standard deviation. The variance measure of BPM may better describe fatigue (Shen, 2018). |
| **Body_Movement** | Average value of the sum of accelerations along the x, y, and z axes (in absolute value) measured every minute through the smartwatch. Body movement is a proxy of the activation level of employees, which in turn affects performance (Gardner, 1990). |
| **Body_Movement_Max** | Maximum value of the variable "Body Movement" in one hour. Body movement peaks can be associated to changes in performance (Gardner, 1990). |
| **Body_Movement_Min** | Minimum value of the variable "Body Movement" in one hour. Also low body movement may be associated to performance variations (Gardner, 1990). |
| **Steps** | Average number of steps per minute in one hour. This was measured through the smartwatch, and it is another measurement of body movement. |

| **Min_Steps** | Minimum value of the variable "Steps" in one hour. |
| --- | --- |
| **N_Steps_Max** | Number of times (in one hour) in which the variable "Steps" exceeds the average "Steps" value by at least one standard deviation. |
| **Teamleader_Int** | Number of interactions between an operator and the team leader in one hour. This measure is obtained considering the proximity of smartwatches. Interactions with the team leader proved to affect both operational and team performance (Easton & Rosenzweig, 2015). |
| **Team Leader Presence** | This value is 1 if the team leader is working and 0 if he is not at work. |

Lagged variables (one-hour) are identified by the prefix "Lag1_". We considered lags of all the above-mentioned measures.

The heart rate-related measurements were normalized based on the individual heart rate average, since this parameter tends to depend on age, weight, and personal characteristics (Jensen-Urstad et al., 1997).

## 4. Results

A machine learning model, namely CatBoost (Prokhorenkova et al., 2018), was trained to assess the predictive power of human and environmental variables on productivity. Results were evaluated through Monte Carlo cross-validation (Dubitzky et al., 2007) with 500 repetitions, i.e., random splits of the dataset into training and test data. On average, our models had a Mean Absolute Error (MAE) of 6.69 and a Root Mean Squared Error (RMSE) of 8.71 on test sets. These results represent a significant reduction of prediction errors with respect to the use of naïve forecasts (MAE = 8.88, RMSE = 11.35), i.e. models which make predictions taking the mean of productivity. In particular, MAE was reduced by 24.7.0% and RMSE by 23.2%. However, our main goal was not to make exact forecasts, but to discover the most significant factors affecting predicted performance.

Subsequently, we applied the average model resulting from the Monte Carlo cross-validation to the full data, in order to evaluate the importance of each predictor, calculated as the mean of its absolute Shapley values (Lundberg & Lee, 2017; Lundberg et al., 2018; Lundberg et al., 2020). These values are shown in Table 2 and Figure 2: the higher the positions of a variable in the chart, the higher its impact on model predictions. All the data analyses were carried out using the Python programming language, specifically the packages SHAP (Lundberg & Lee, 2017) and CatBoost (Prokhorenkova et al., 2018).

**Table 2** - Mean Absolute Shapley Values

| Variable | M | SD | Variable | M | SD |
|---|---|---|---|---|---|
| USERID | 3.75 | 1.73 | HUMIDITY | 0.20 | 0.24 |
| INT_USER728 | 1.10 | 1.31 | Lag1_STEPS_MIN | 0.20 | 0.17 |
| BPM | 0.67 | 0.41 | BODY_MOVEMENT_MAX | 0.19 | 0.20 |
| Lag1_TEAM_LEADER_INT | 0.51 | 0.15 | Lag1_OPERATORS_INT | 0.19 | 0.41 |
| Lag1_INT_USER728 | 0.48 | 0.17 | Lag1_N_STEPS_MAX | 0.19 | 0.15 |
| BPM_MAX | 0.43 | 0.41 | INT_USER726 | 0.17 | 0.39 |
| DAY | 0.40 | 0.39 | STEPS_MAX | 0.17 | 0.19 |
| LUMINOSITY | 0.40 | 0.36 | Lag1_INT_USER726 | 0.16 | 0.15 |
| Lag1_STEPS | 0.38 | 0.29 | TEAM_LEADER_PRESENCE | 0.16 | 0.12 |
| Lag1_BPM | 0.37 | 0.29 | Lag1_BODY_MOVEMENT_MIN | 0.15 | 0.13 |
| STEPS_MIN | 0.36 | 0.22 | OPERATORS_INT | 0.15 | 0.17 |
| WEEK | 0.34 | 0.35 | WORKED_HOURS | 0.14 | 0.22 |
| MICROPHONE_MAX | 0.33 | 0.27 | N_STEPS_MAX | 0.14 | 0.14 |
| INT_USER727 | 0.32 | 0.36 | Lag1_BODY_MOVEMENT | 0.13 | 0.09 |
| TEAM_LEADER_INT | 0.31 | 0.23 | Lag1_MICROPHONE | 0.12 | 0.14 |
| TEMPERATURE | 0.29 | 0.41 | Lag1_MICROPHONE_MAX | 0.12 | 0.12 |
| HOUR | 0.27 | 0.21 | Lag1_STEPS_MAX | 0.11 | 0.12 |
| BODY_MOVEMENT_MIN | 0.27 | 0.30 | MICROPHONE | 0.10 | 0.13 |
| Lag1_INT_USER731 | 0.26 | 0.25 | N_BPM_MAX | 0.08 | 0.09 |
| BODY_MOVEMENT | 0.25 | 0.23 | Lag1_TEMPERATURE | 0.07 | 0.11 |
| STEPS | 0.23 | 0.35 | Lag1_INT_USER727 | 0.04 | 0.06 |
| Lag1_LUMINOSITY | 0.22 | 0.17 | Lag1_N_BPM_MAX | 0.04 | 0.06 |
| INT_USER731 | 0.20 | 0.16 | Lag1_HUMIDITY | 0.00 | 0.00 |
| Lag1_BPM_MAX | 0.20 | 0.18 | Lag1_TEAM_LEADER_PRESENCE | 0.00 | 0.00 |
| AFTERNOON | 0.20 | 0.19 | Lag1_BODY_MOVEMENT_MAX | 0.00 | 0.00 |

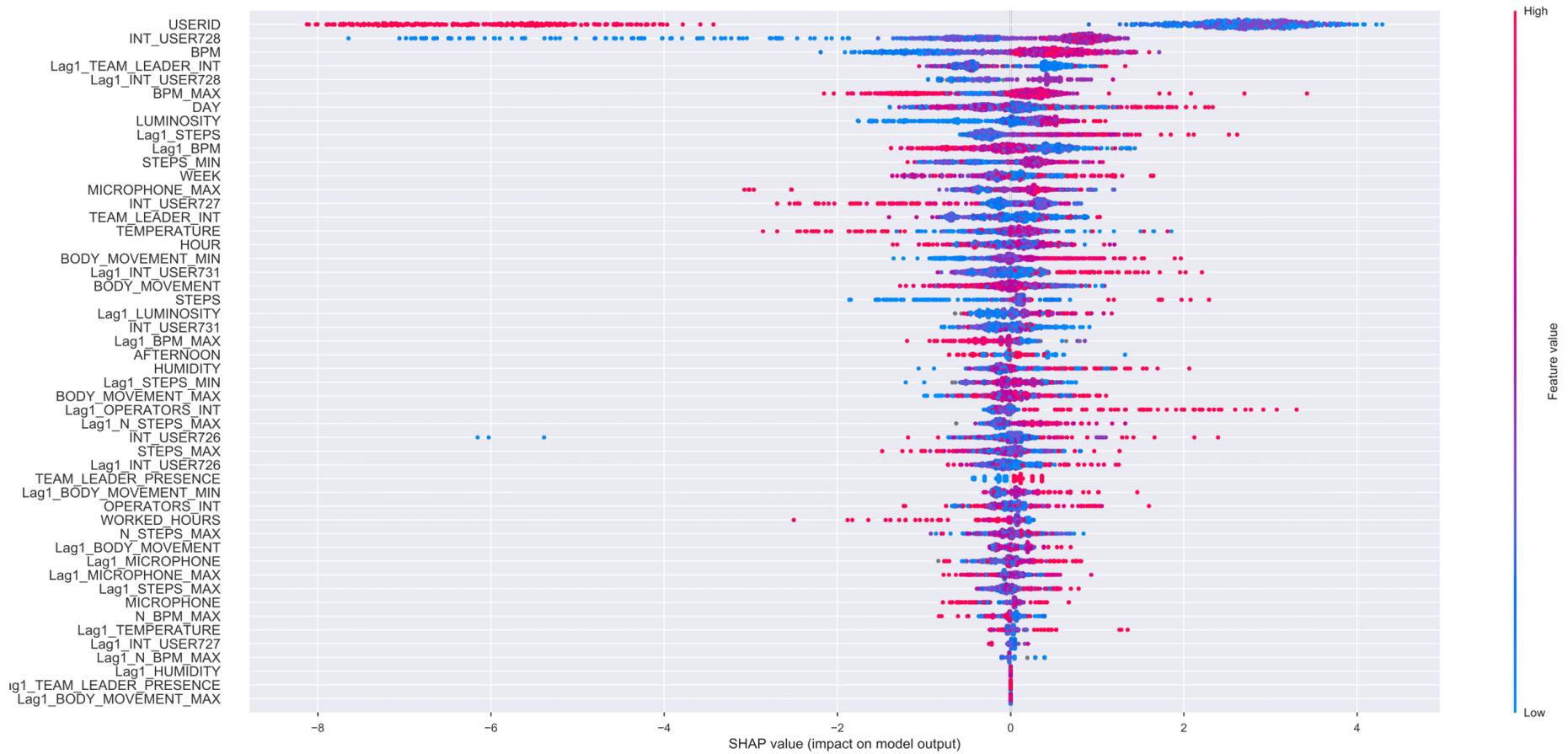

**Figure 2** - Feature Importance

Features with SHAP values more distant from the vertical axis (zero value) have a higher influence on the model predictions. As Figure 2 and Table 2 show, UserID is the most important control variable (with the operator #731 being the least productive). This is not a surprise and reflects the idea that productivity depends on people, with some individuals being way more productive than others.

Consistently, it is interesting to notice that the level of interaction with some individuals seems to have more impact on the productivity than others. Indeed, the level of interaction with the operator #728 is the second most important predictor in the model, with low values suggesting a lower performance of the other employees. Having a high level of interaction with the operator #727, on the other hand, seems to be detrimental to individual performance – whereas interacting with the operators #731 and #726 is of little impact.

Figure 3 shows the SHAP values at different levels of interaction with Users #727 and #728 (variables INT_USER728 and INT_USER727).

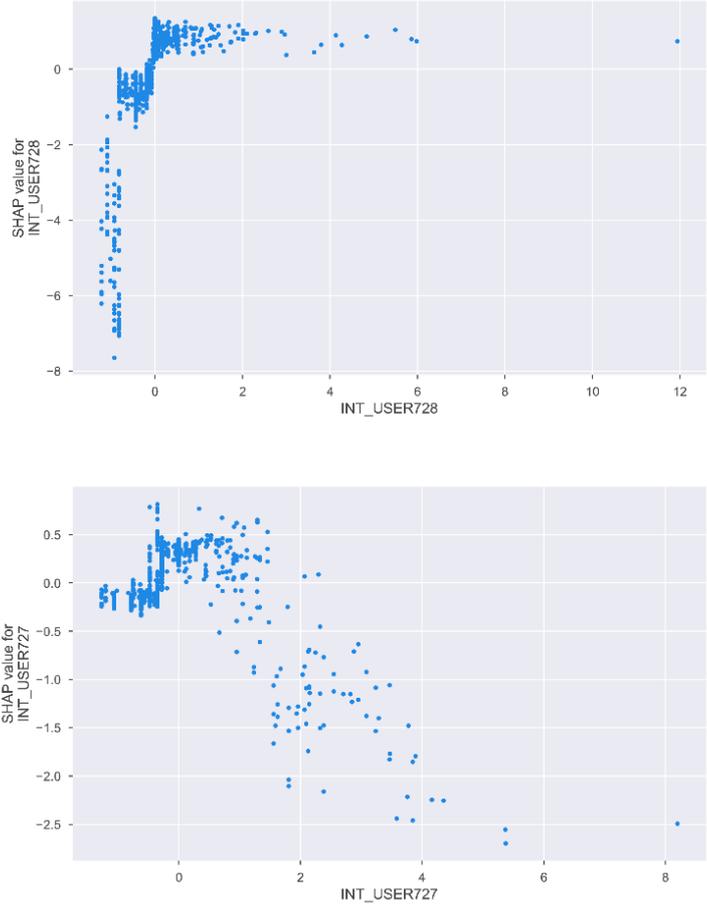

**Figure 3** - Interactions with operators #727 and #728

Looking at the plots (Figure 3), we see that the lower the values of interaction with operator #728 are, the lower are the SHAP values (i.e., more probable predictions of low productivity). This effect is particularly pronounced for the interaction values below the average, while it tends to wane for high levels of interaction. By contrast, interacting with operator #727 seem to be positive for productivity only up to a certain level, with small or positive impact up to 1.5 standard deviations. However, when the value of interaction with operator #727 (INT_USER727) gets higher, the predicted performance is much lower.

Figure 4 shows the effect of interacting with the team leader. Observations in the plot (Figure 4) are colored by UserID, to show how the effect changes for different employees. Indeed, the impact of team leader interaction (TEAM_LEADER_INT) highly depends on the operator – with some employees taking more advantage of the team leader supervision than others. For example, operator #731 has significantly higher hourly productivity when his interactions with the team leader are low.

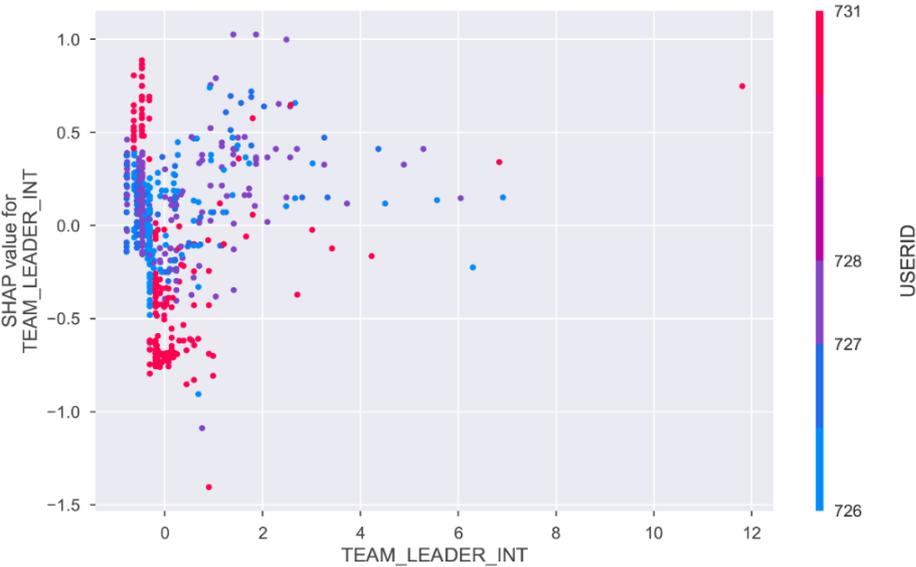

**Figure 4** - Team leader interactions

In these models, we also controlled for the effect of time and working shifts (afternoon vs morning) on productivity. It is no surprising that productivity of employees decreases in the second half of the shift (after the 4$^{th}$ hour), with the steepest declining trend in the last two hours.

We also discovered that the effect of interaction on performance is dependent on the time of the day: a higher level of interaction seems to be beneficial in the morning, whereas many interactions in the afternoon seem to lower productivity (Figure 5). Similarly, we notice that some operators were considerably more productive during specific days of the observation. Operator #731 is confirmed to be the most susceptible, presenting greater variability than the other workers.

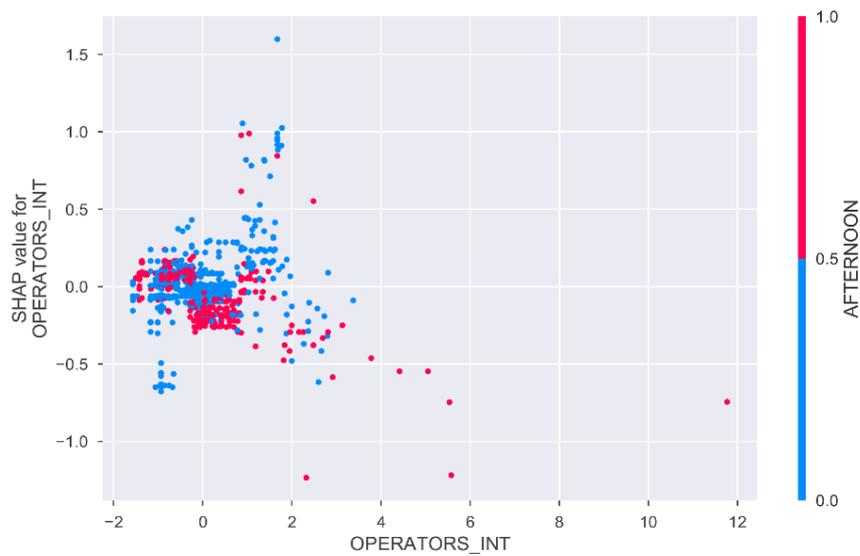

**Figure 5** - Working shift and Operators' interactions

Digging deeper into the information provided by the analysis of body signals, we found that BPM is the third most important predictor. The SHAP values of this variable increase almost linearly up to one standard deviation above the (individual) average (Figure 6). On the other hand, when BPM is too high, the productivity tends to slightly shrink. Similar evidence also emerges from the analysis of heart rate peaks, which are usually associated with lower predicted performance – they seem to represent cases of overexertion, or altered emotional states. Similarly to BPM, productivity seems to be greater when the minimum value of body movement is higher (Body_Movement_Min), as also shown in Figure 6.

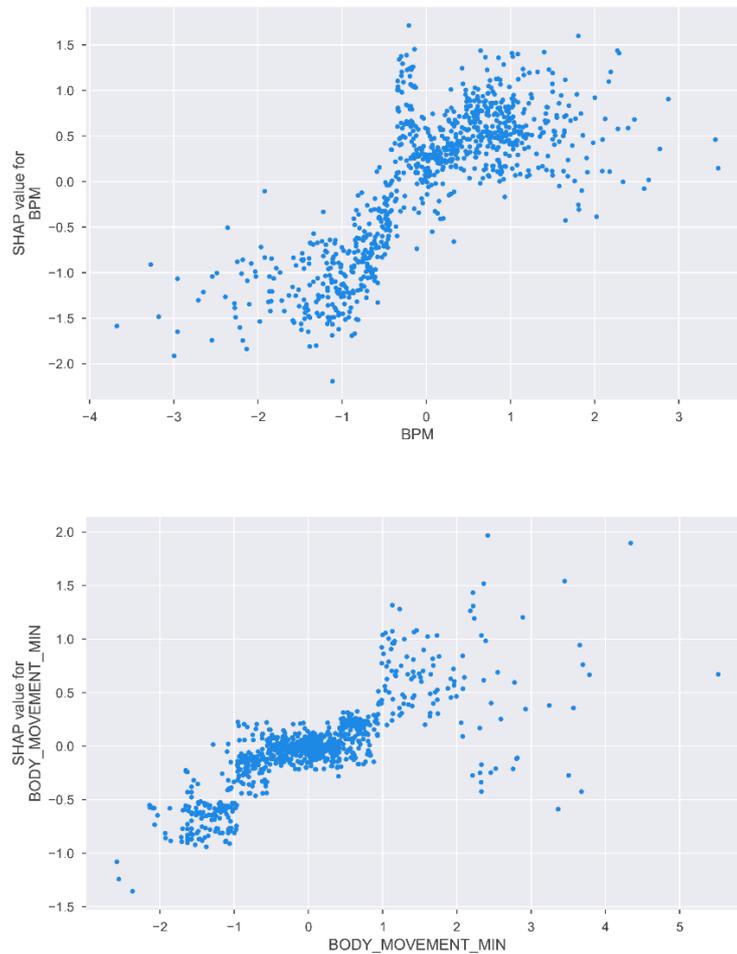

**Figure 6** - BPM and Body Movement

However, the positive effect of this variable appears to be significant up to two standard deviations above the average.

Finally, environmental conditions appear to have a relevant impact on productivity. Specifically, luminosity has a positive effect on performance: the more intense the illumination, the higher the predicted productivity (Figure 7). However, this relationship is not linear because the negative effect of low values of luminosity is particularly strong. The size of this effect also changes depending on the individual, with some employees being more affected by this environmental condition than others. For example, operator #731 seems to be more sensitive to illumination, demonstrating increased performance in conditions of bright lighting.

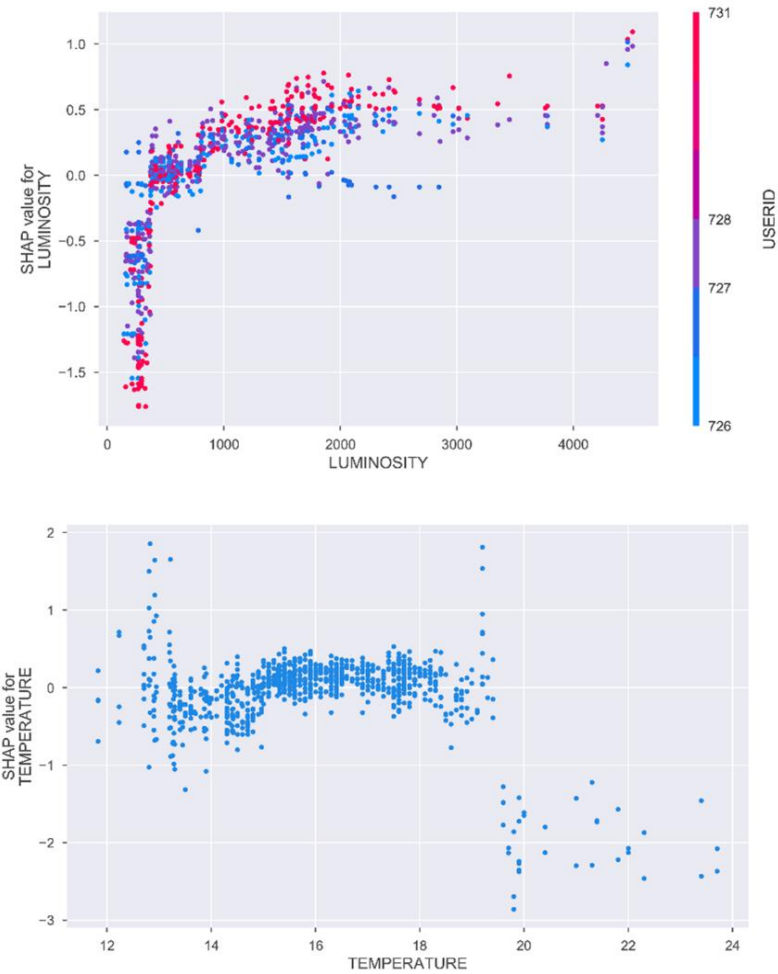

**Figure 7** - Luminosity and Temperature

Humidity and temperature seem not to affect predictions, except when they are particularly low or high. In particular, high temperature appears to predict lower performance, though the chosen test period – i.e. early spring – rarely presented such condition (Figure 7).

## 5. Discussion and Managerial Implications

### *5.1 Theoretical contribution*

Consistently with past research analyzing different human-intensive operational contexts (e.g. Loch & Wu, 2007; Donohue et al., 2018; Fahimnia et al., 2019; Stefanini et al., 2020), our findings suggest that human and environmental factors play a relevant role in logistics activities, significantly affecting system performance. The results obtained through our case study permitted to explore such – often neglected – factors, trying to respond to the first RQ.

Specifically, workers' performance appears to be highly influenced by individual characteristics, with some operators being more productive than others while working under the same conditions (De Vries et al., 2016; Obembe, 2010; Othman & Mahmood, 2019). This finding supports the need for an individualized resource management and the idea that no standard rules – "golden rules" – can be applied to all settings and employees. Employees are also differently influenced by boundary conditions, with some operators more susceptible to the contingent context than others. Despite such a result seeming straightforward, the attention paid to such variables is scarce, with both researchers and managers considering mainly system design aspects of logistics hubs such as routing, layouts, and travel distances (Grosse et al., 2017; Masae et al., 2020; Katiraee et al., 2021). Neglecting individual features in the study and design of logistics activities, as well as intensive operational ones, may mean missing important determinants of system performance. Our methodology allowed us to understand which specific work conditions enable better performance of each individual and/or group – supporting the assessment of the effect of managerial practices, or other events, that take place on specific days.

Our results also assert the relevance of interactions between workers. Specifically, some workers seem to have a positive influence on the hourly productivity of colleagues. This effect may be due to positive problem-solving cooperation with more skilled or experienced employees. This might include for instance asking a colleague how to fix a problem with the picking scanner, or where specific references are located, or how to effectively handle the material. On the contrary, excessive interaction with some operators appears to be detrimental to the productivity of colleagues. This seems to happen especially when tiredness is high (e.g. in the afternoon), thus it is probably due to distractions from working activities. Such findings provide first empirical evidence about the importance of workers' collaboration dynamics in logistics, as theoretically and empirically found in other operational environments by past literature (e.g. Mathieu et al., 2008; Grosse et al., 2015; Forsyth, 2018; Bellingan et al., 2020).

Another important factor for individual productivity in logistics activities seems to be the relationship of employees with their supervisor, represented here by the team leader. Some operators tend to have higher productivity when interacting more with the team leader, while others have a significantly higher productivity when such interactions are low. This might depend on the different attitudes of employees, their skills, capabilities and experience in managing operational tasks, collaborating with colleagues and dealing with stressful conditions – e.g. the "stress of control" and "pressure on results" that the closeness of a supervisor can exert (Belbin, 2012; Hackney et al., 2018). For this reason, consistent with management literature (e.g. Ellinger et al., 2005; Cheung & Wong, 2011, Kafetsios et al., 2011; Chemers, 2014), logistics managers should adapt their style of leadership to take in account the different psychological characteristics of each collaborator.

Individual performance appears also to be affected by physical conditions, i.e. the level of "activation" of operators. In particular, low heart rate and limited body movement are associated with lower levels of productivity. On the contrary, when employees are physically active, they are more ready and more prone to work, and they tend to perform their tasks better and faster. This insight is also supported by the drop in performance depending on the shift, with the last hours being less productive than the first ones.

However, very high peaks of heart rate and body movement can be related to lower performance because such features are linked to an agitated state of workers (e.g. anxiety, nervousness, distress, anger). This result, aligned with past literature (Winkelhaus *et al.* (2018), Makkar &

Basu, 2019; Stefanini et al., 2020), suggests that fatigue, altered emotional states, and overstressed working conditions lead individuals to worse performance.

Our findings also reveal a potentially remarkable role of environmental factors during logistics activities, as conceptually sustained by past literature (Danna & Griffin, 1999; Grosse et al., 2017). Specifically, we identified a possible influence of luminosity on operators' performance (Füchtenhans et al., 2021). As expected, more intense illumination seems to increase the productivity of workers. This relationship is especially significant for very low values of luminosity, where the operators' performance tends to drop markedly. This effect can have two explanations: 1) the light stimulates operators to be more active and work harder; 2) darkness (lower luminosity) makes it more difficult for workers to operate with forklifts and shuttles inside the warehouse. However, the size of the positive influence of luminosity varies on an individual basis. Therefore, results not only attribute importance to environmental factors, but they also show that employees can have very different sensitivities to environmental conditions.

In addition, although the observations were done in early spring and were barely affected by high temperatures, findings appear to suggest a negative impact of high temperatures on performance. This effect is likely due to the discomfort of operators caused by heat.

From a methodological perspective, this study provides a first answer to the second RQ, addressing the issue of "how to" quantitatively investigate behavioral factors in operations management (Croson et al., 2013; Bendoly et al., 2015; Brocklesby, 2016). Indeed, this research represents an early attempt to investigate the effect of human and environmental aspects on performance in logistics, using direct and quantitative measurements of workers' behavior and workplace conditions. The simultaneous usage of wearable sensors, thermo-hygrometer and lux-meter with the use of a machine learning model to exploit collected data is easily reproducible in other working contexts or for further studies. Moreover, it offers a way to reduce complexity in the analysis of multiple interacting factors that can impact performance (Van Knippenberg & Schippers, 2007) – and were rarely considered altogether in past research. Our findings are promising and encourage further similar studies exploring human and environmental aspects in logistics and other human-intensive activities.

Exploiting innovative smart tools and a novel machine learning method, our study also represents an empirical contribution to the call for investigations on the implications of Big Data and innovative data science methods for management research (Dubey et al., 2019). Indeed, the application of novel approaches for organizational studies in real business settings, based on new tools (e.g., wearable sensors, social network platforms, smartphones) and methods of analysis (e.g., data mining, machine learning), is challenging and still quite limited albeit it is very promising (George et al., 2016; Chaffin et al., 2017; Ni et al., 2020). As shown by our results, such an analysis can provide new opportunities to evaluate neglected aspects of operations management (George et al., 2016).

### 5.2 *Managerial contribution and practical implications*

This study offers managerial insights on how human and environmental factors can affect logistics performance. Specifically, the managerial contributions are twofold: on the one hand, the study provides an innovative approach to monitor the behaviors and well-being of employees and the dynamics of collaboration among them; on the other hand, it presents a series of specific variables, both individual and environmental, that management can consider to design and to control material handling processes in logistics hubs.

Accordingly, the first advice for managers is to take into consideration the individual characteristics of employees in a "customize management" perspective (Burinskiene, 2010; De Vries et al., 2016; Gawron, 2016; Obembe, 2010; Othman & Mahmood, 2019). Logistics workers are usually considered fully interchangeable, similarly to robots, but individual features matter as proven by our findings, therefore such assumptions are often false. Understanding the characteristics and behaviors of employees (i.e., how different workers react to different conditions) may allow managers to profile them on an individual basis – while fully respecting individual privacy – and be able to better assist them to reach their full potential, for example designing a proper work environment and improving working group composition. In addition, some work conditions could be customized on an individual basis, taking in account the emotional component of more susceptible workers, to allow them to increase their performance. Lastly, individual psychological characteristics should be better considered when hiring new employees for logistics hubs. Moreover, logistics managers should pay adequate attention to the interactions among employees during working shifts. They may favor appropriate collaboration dynamics by offering incentives and rewards based on the productivity of the team and by promoting healthy competition between different teams. This would encourage the least productive employees to work more productively and would make workers more willing to help team members, keeping in mind that workers' motivation and well-being also depend on the relationship with colleagues (Bengtsson & Kock, 2014; Bouncken et al., 2015; Jordan et al., 2019; Bellingan et al., 2020).

The team leader also plays an important role in collaboration dynamics (Boyatzis et al., 2008; Hirst et al., 2016). He should support workers throughout the work, fixing issues, enhancing their well-being, and facilitating team cooperation. In addition, supervisors should consider to lead operators differently, taking into account the diverse psychological characteristics of each. Our approach offers the possibility to train team leaders, by informing them about the outcomes of their (inter)actions – i.e., providing a mirror on the effects of their behaviors, which ultimately triggers organizational change (Gloor et al., 2017; Hattie & Timperley, 2007; Ramos, 2007).

Other fundamental aspects to be monitored by managers are related to workers well-being and environmental conditions. Specifically, an adequate illumination in the warehouse is desirable to put operators in the right condition for material handling activities and to keep them more "active". Finally, managers should try to maintain a pleasant temperature in the workplace, avoiding high and low extremes.

Although it is challenging to control for all the determinants of logistics performance linked to human and environmental factors, these recommendations may help managers in planning logistics activities, designing warehouse structures, hiring staff and improving leadership, coordination, and collaboration skills.

Contributions of this study include a potential positive impact for society, particularly referring to the social development goal (SDG #8). Indeed, this work promotes the awareness and the improvement of employees' well-being, even if benefits are yet to be proven by practical adoptions of such a methodology and operators might not be willing to wear such devices.

## 6. Conclusions

### 6.1 General Conclusions

This research empirically explores human and environmental factors affecting the performance of workers during logistics activities by using a novel approach based on the use of body sensors, smart sensors, interaction analysis and machine learning. Specifically, the research gaps highlighted by the two RQs have been addressed thanks to this study.
Findings suggest that "people", through their individuality, behavior, and collaboration dynamics remarkably influence the performance of logistics activities. In addition, environmental conditions of the workplace seem to be relevant for workers' productivity, and their impact may vary depending on individual characteristics and sensitivity. Although such relationships are theoretically predicted by past research (e.g., Grosse et al., 2015), empirical studies considering the impact on logistics operations of both human and working environmental factors are almost absent (Grosse et al., 2017; Masae et al., 2020). However, as shown by our findings, neglecting such factors means relinquishing a holistic, effective, and factual investigation of work dynamics in logistics systems. Accordingly, the results obtained also offers relevant managerial and practical implications as discussed in Section 5.2.

### 6.2 Limitations and Future Research

The most important limitations of this study are related to the exploratory nature of the work. Drawing on a single case study, results might be affected by the specific application context. This is a common issue for many behavioral studies that limits the generalization of results (Tröster et al., 2014). Moreover, even though the number of monitored variables and related indicators was quite high, additions would be possible. Many other potentially significant metrics describing human and environmental factors may deserve to be included, such as for example the mood of employees or their personality traits. Observations were carried out during the early spring, thus the most extreme climate conditions (e.g., very hot and humid, or very cold weather) were likely excluded from our sample. Finally, the number of employees involved in the exploratory study is quite low.

Despite these limitations the presented research presents a novel perspective on the quantitative analysis of human and environmental impacts on workers' productivity. This can serve as a starting point for a more in-depth analysis aimed at overcoming the limitations of this work.

This study's limitations point out some directions for future research. Future studies may reinforce our preliminary evidence by extending the observation period by including different seasons and by considering a higher number of employees. Future research could also replicate the study in different warehouse contexts, because a different cultural context, automation level and number of workers can significantly affect the operating scenario. The inclusion of a larger number of environmental sensors is also desirable to get more detailed insights on the role of environmental aspects. Another extension could be made by integrating personality aspects of workers with their physical factors, providing a more complete definition of individuality.

The conjoint application of this approach of analysis with other digital technologies may also be promising. For example, future implementations of AR or VR technologies may integrate the system developed in this study to evaluate the AR/VR effects on the individual and collaborative behaviors of workers and on their performances.

Lastly, our methodology could be applied in different manufacturing and service contexts to confirm its suitability for investigating human and environmental factors in the operations management field.